\message{Uncomment writedefs and}
\message{TEX TWICE TO GET LABELS RIGHT!!!}

\input harvmac
\input labeldefs.tmp


\def\eql{~=~}

\def\al{\alpha}
\def\ga{\gamma}
\def\Om{\Omega}

 \def\cI{{\cal I}}
 
\def\cL{{\cal L}} 
\def\cN{{\cal N}} \def\cO{{\cal O}}
 
\def\cR{{\cal R}} 
 
\def\cZ{{\cal Z}}
\def\ph{\phi}
\def\om{\omega}
\def\de{\delta}
\def\ie{{\it i.e.\/}}
\def\bfone{\relax{\rm 1\kern-.35em 1}}
\def\IC{\relax\,\hbox{$\inbar\kern-.3em{\rm C}$}}
\def\ID{\relax{\rm I\kern-.18em D}}
\def\IF{\relax{\rm I\kern-.18em F}}
\def\IH{\relax{\rm I\kern-.18em H}}
\def\II{\relax{\rm I\kern-.17em I}}
\def\IN{\relax{\rm I\kern-.18em N}}
\def\IP{\relax{\rm I\kern-.18em P}}
\def\IQ{\relax\,\hbox{$\inbar\kern-.3em{\rm Q}$}}
\def\us#1{\underline{#1}}
\def\IR{\relax{\rm I\kern-.18em R}}
\font\cmss=cmss10 \font\cmsss=cmss10 at 7pt
\def\ZZ{\relax\ifmmode\mathchoice
{\hbox{\cmss Z\kern-.4em Z}}{\hbox{\cmss Z\kern-.4em Z}}
{\lower.9pt\hbox{\cmsss Z\kern-.4em Z}}
{\lower1.2pt\hbox{\cmsss Z\kern-.4em Z}}\else{\cmss Z\kern-.4em
Z}\fi}
\def\th{\theta}
\def\ep{\epsilon}
\def\si{\sigma}
\def\De{\Delta}
\def\rh{\rho}
\def\al{\alpha}
\def\be{\beta}
\def\XXone{{X_1}}
\def\XXtwo{{X_2}}
\def\ccr{c}
\def\rhr{\rh}

\def\slashed#1{\ooalign{\hfil\hfil/\hfil\cr $#1$}}

%
\def\nihil#1{{\it #1}}
\def\eprt#1{{\tt #1}}
\def\nup#1({Nucl.\ Phys.\ $\us {B#1}$\ (}
\def\plt#1({Phys.\ Lett.\ $\us  {#1B}$\ (}
\def\cmp#1({Comm.\ Math.\ Phys.\ $\us  {#1}$\ (}
\def\prp#1({Phys.\ Rep.\ $\us  {#1}$\ (}
\def\prl#1({Phys.\ Rev.\ Lett.\ $\us  {#1}$\ (}
\def\prv#1({Phys.\ Rev.\ $\us  {#1}$\ (}
\def\mpl#1({Mod.\ Phys.\ Let.\ $\us  {A#1}$\ (}
\def\ijmp#1({Int.\ J.\ Mod.\ Phys.\ $\us{A#1}$\ (}
\def\jag#1({Jour.\ Alg.\ Geom.\ $\us {#1}$\ (}

\def\cI{1}
%
\lref\MartinecQN{
E.~J.~Martinec,
``Integrable Structures in Supersymmetric Gauge and String Theory,''
Phys.\ Lett.\ B {\bf 367}, 91 (1996)
[arXiv:hep-th/9510204].
}
\lref\DonagiCF{
R.~Donagi and E.~Witten,
``Supersymmetric Yang-Mills Theory And Integrable Systems,''
Nucl.\ Phys.\ B {\bf 460}, 299 (1996)
[arXiv:hep-th/9510101].
}
%
\lref\PilchUE{
K.~Pilch and N.~P.~Warner,
``N = 2 supersymmetric RG flows and the IIB dilaton,''
Nucl.\ Phys.\ B {\bf 594}, 209 (2001)
[arXiv:hep-th/0004063].
}
%
\lref\BrandhuberCT{
A.~Brandhuber and K.~Sfetsos,
``An N = 2 gauge theory and its supergravity dual,''
Phys.\ Lett.\ B {\bf 488}, 373 (2000)
[arXiv:hep-th/0004148].
}
\lref\BuchelCN{
A.~Buchel, A.~W.~Peet and J.~Polchinski,
\nihil{Gauge dual and noncommutative extension of an N = 2 supergravity  
solution}, Phys.\ Rev.\ D {\bf 63} (2001) 044009, 
\eprt{hep-th/0008076}.}
%
\lref\EvansCT{
N.~Evans, C.~V.~Johnson and M.~Petrini,
\nihil{The enhancon and N = 2 gauge theory/gravity RG flows},
JHEP {\bf 0010} (2000) 022, 
\eprt{hep-th/0008081}.}
%
\lref\SchwarzQR{
J.~H.~Schwarz,
\nihil{Covariant Field Equations Of Chiral N=2 D = 10 Supergravity},
Nucl.\ Phys.\ B {\bf 226} (1983) 269.
}
%
\lref\GauntlettSC{
J.~P.~Gauntlett, D.~Martelli, S.~Pakis and D.~Waldram,
``G-structures and wrapped NS5-branes,''
arXiv:hep-th/0205050.
}
%
\lref\GauntlettNW{
J.~P.~Gauntlett, J.~B.~Gutowski, C.~M.~Hull, S.~Pakis and H.~S.~Reall,
``All supersymmetric solutions of minimal supergravity in five  dimensions,''
arXiv:hep-th/0209114.
}
%
\lref\GauntlettFZ{
J.~P.~Gauntlett and S.~Pakis,
``The geometry of D = 11 Killing spinors,''
arXiv:hep-th/0212008.
}
%
\lref\PopeJP{
C.~N.~Pope and N.~P.~Warner,
``A dielectric flow solution with maximal supersymmetry,''
arXiv:hep-th/0304132.
}
\lref\GNW{
C.~Gowdigere, D.~Nemeschansky and N.~P.~Warner,
``Supersymmetric Solutions with Fluxes from Algebraic Killing Spinors,'' 
USC-03/02,  arXiv:hep-th/xxxxxxx . }
%
\lref\GowdigereUK{
C.~N.~Gowdigere and N.~P.~Warner,
``Flowing with eight supersymmetries in M-theory and F-theory,''
arXiv:hep-th/0212190.
}
%
\lref\BuchelAH{
A.~Buchel and J.~T.~Liu,
arXiv:hep-th/0305064.
}
%
\lref\BuchelQM{
A.~Buchel,
``Compactifications of the N = 2* flow,''
arXiv:hep-th/0302107.
} 
%
%
\lref\LuTH{
H.~Lu, C.~N.~Pope and J.~Rahmfeld,
``Killing spinors on spheres, AdS and hyperbolic spaces,''
{\it Prepared for Richard Arnowitt Fest: A Symposium on Supersymmetry and 
Gravitation, College Station, Texas, 5-7 Apr 1998}
}
%
\lref\LuNU{
H.~Lu, C.~N.~Pope and J.~Rahmfeld,
``A construction of Killing spinors on S**n,''
J.\ Math.\ Phys.\  {\bf 40}, 4518 (1999)
[arXiv:hep-th/9805151].
}
%

\Title{
\vbox{
\hbox{USC-03/05}
\hbox{\tt hep-th/0306098}
}}
{\vbox{\vskip -1.0cm
\centerline{\hbox{Generalizing the $N=2$ supersymmetric RG flow }}
\vskip 8 pt
\centerline{\hbox{solution of IIB supergravity}}}}
\vskip -.3cm
\centerline{Krzysztof Pilch and Nicholas P.\ Warner }
\medskip
\centerline{{\it Department of Physics and Astronomy}}
\centerline{{\it University of Southern California}}
\centerline{{\it Los Angeles, CA 90089-0484, USA}}

\bigskip
\bigskip
We explicitly construct the supersymmetry transformations for the
$\cN=2$ supersymmetric RG flow solution of chiral IIB supergravity. We
show that the metric, dilaton/axion, five-index tensor and half of the
three index tensor are determined algebraically in terms of the
Killing spinor of the unbroken supersymmetry.  The algebraic nature of
the solution allows us to generalize this construction to a new class
of $\cN=2$ supersymmetric solutions of IIB supergravity. Each solution
in this class is algebraically determined by supersymmetry and is
parametrized by a single function of two variables that satisfies a
non-linear equation akin to the Laplace equation on the space
transverse to the brane.

\vskip .3in
\Date{\sl {June, 2003}}

\vfill\eject

\newsec{Introduction}

There continues to be a strong interest in finding supersymmetric
backgrounds with non-trivial $RR$-fluxes.  Indeed, one of the most
interesting forms of this problem is to start with a geometric
background with a higher level of supersymmetry and then use the
fluxes to break some, or all of this supersymmetry.  This has been a
theme of much recent research, but here we will focus on its
application in holographic field theory, and most particularly in the
four-dimensional AdS/CFT correspondence.  The problem is then one of
finding supersymmetric solutions of IIB supergravity in which there
are non-trivial fluxes that vanish suitably at infinity.  To
be more specific, this paper will examine the problem of holographic
flows of $\cN=4$, $SU(N)$ Yang-Mills theory in which the supersymmetry
is softly broken to $\cN=2$ (eight supersymmetries).

Such flows are well-understood in field theory, and the Wilsonian
effective action has been computed in \refs{\MartinecQN,\DonagiCF}.
They thus provide a valuable test of the holographic correspondence.
These theories have a Coulomb branch parametrized by the invariants of
the scalar field, $\Phi$, that lies in the $\cN=2$ vector multiplet:
\eqn\Cinvs{ u_m ~\equiv~ {\rm Tr}\,\big( \Phi^m \big)\,, 
\qquad m=1,2, \dots N \,,}
and in the large-$N$ limit there is thus an infinite series of such
invariants.  In terms of branes, this Coulomb branch involves the $D3$
branes spreading out in the two directions, $(v_1,v_2)$, that
correspond to (complex) vevs of $\Phi$.  The $u_m$ are then the moments of the
brane distribution.  For large $N$, a general point on the Coulomb branch therefore
corresponds to an arbitrary function, $\rho(v_1,v_2)$, that defines
the density of the $D3$-brane distribution in the two-dimensions in
which the spreading occurs.  If the supersymmetry were not softly
broken then a solution with maximal supersymmetry could be obtained
via the usual ``harmonic rule,'' with the harmonic function sourced by
the distribution, $\rho(v_1,v_2)$.  The corresponding solution with
soft supersymmetry breaking to $\cN=2$ is unknown for general
$\rho(v_1,v_2)$, and indeed this has been an open problem for several
years.

The solution for one particular, relatively uniform distribution of branes was obtained
in \PilchUE, and the precise brane distribution was computed in \refs{\BuchelCN, \EvansCT}.   
This solution has found, and continues to find (see, for example, \refs{\BuchelAH,\BuchelQM})  
interesting applications,  and it would be rather useful to  find $\cN=2$
supersymmetric holographic flows to a broader section
of the Coulomb branch.   The essential problem is the apparent complexity
of the solution of \PilchUE.   In this paper we re-examine the solution of
\PilchUE\ and explicitly compute the Killing spinors.  We find that while the
geometry is complicated, the structure of the Killing spinors is remarkably
simple:  The Killing spinors are determined {\it algebraically} in terms of
the metric coefficients. Conversely, we find that much of the geometry and the fluxes 
can the be inferred from these Killing spinors.  This enables us to make a more general
Ansatz that will capture rotationally symmetric distributions of $D3$ branes, that is,
density functions, $\rho(v_1,v_2)$, that only depend upon 
$v \equiv  \sqrt{v_1^2 + v_2^2}$.

Our approach to finding the softly broken flows is thus parallel
to that of \GNW:  We make a very general Ansatz
for the metric and for the fluxes, and then impose projectors that determine
the Killing spinor spaces algebraically in terms of the metric coefficients.  There
is an invaluable constraint on the Killing spinors:  Namely, if $\epsilon^a$ and
$\epsilon^b$ are two commuting (non-grassmann) Killing spinors, then
\eqn\KillStr{K_{(ab)}^\mu ~\equiv~ \overline{\epsilon^{a} } 
\gamma^\mu \epsilon^{b}~+~ \overline{\epsilon^{b} } 
\gamma^\mu \epsilon^{a}\,,}
must be a Killing vector of the underlying background metric.   We will
refer to this linked property of the Killing vectors and spinors as
the ``Killing Structure.''  This structure provides strong constraints
on the projectors that define the Killing spinors, and also enables us
to completely fix the normalizations of the Killing spinors.  Thus the
simple structure of the flow solutions lies in the definition of the
Killing spinors.  We then use the supersymmetry transformations to
reconstruct everything else.  These transformations can be used to fix
the fluxes and metric algebraically in terms of the functions (and
their derivatives) that define the Killing spinors.  Finally, the
Bianchi identities provide the necessary differential equations.
Indeed, we find that metric and all the fluxes are can be
algebraically determined in terms of a single function, $c$, and its
derivatives.  The function, $c$, must itself satisfy a second order,
partial differential equation.  The only difficulty is that this
differential equation is non-linear, but like the corresponding
$M$-theory result \GNW, this differential equation has a rather
straightforward perturbation theory that shows that it does indeed
admit solutions with a functional degree of freedom corresponding to a
general, rotationally symmetric, two-dimensional brane distribution,
$\rho(v)$.  Thus, while the governing differential equation is
non-linear, we have found a natural generalization of the ``harmonic
rule'' to softly broken supersymmetry.

In the next section of   we summarize the key properties of the solution of
\PilchUE, and in Section 3 we compute the Killing spinors of the supersymmetry
explicitly.  In particular, we show how the space of Killing spinors
can be defined by two projectors, $\Pi^{(j)}, j =1,2$, that are
algebraically related to the metric coefficients.  Each of these
projectors reduces the dimension of the spinor space by a factor of
two, and together they define the space of the eight supersymmetries.
One of the projectors is naturally interpreted in terms of a helicity
projector on the moduli space, $(v_1,v_2)$, of the $D3$-branes, while
the other projector is a ``dielectric deformation'' of the usual
projector ($\half(1 -i\gamma^{1234})$) associated with a stack of
$D3$-branes.  The Killing structure helps fix this deformation, and
determines the normalization of the Killing spinors.  In Section 4 we
use these observations to obtain a more general Ansatz for $\cN=2$
flows with a rotationally symmetric distribution of branes, and we
then solve this Ansatz and show that the entire solution is generated
by the single function, $c$, that satisfies a particular second order,
non-linear PDE.  We then find some simple solutions of this PDE and
use them to generate new backgrounds with eight supersymmetries.
Section 5 contains some final remarks.

\newsec{The $\cN=2$ supersymmetric RG flow solution}
\seclab\rgflowsolution

In this section we summarize the details of the solution of the chiral
IIB supergravity obtained in \PilchUE.  This solution 
corresponds to an $\cN=2$ supersymmetric RG flow of the $\cN=4$
super-Yang-Mills, and was obtained by lifting to ten
dimensions a solution of $\cN=8$ gauged supergravity in five
dimensions.  In such a construction it is natural to use
coordinates in ten dimensions in which both the flow and the lift are
manifest, even though, as will become clear later, the ten-dimensional
geometry may be somewhat obscured. For the moment we will follow the
convention in \refs{\PilchUE,\SchwarzQR} so that a comparison with the result there 
is more straightforward. In particular, $x^\mu$, $\mu=0,\ldots,3$ are
coordinates along the brane, $r$ is the coordinate along the flow,
while $\th$, $\al^i$ and $\ph$ are coordinates on the deformed $S^5$,
where $\al^i$, $i=1,2,3$, are the Euler angles of the unbroken $SU(2)$.
We will denote the ten-dimensional coordinates collectively by $x^M$,
where $M$ runs from 1 to 10.  Following \refs{\PilchUE,\SchwarzQR}, our
metrics will be ``mostly minus.''

The solution involves two functions, $c(r)$ and $\rh(r)$, that are
related to the mass and the Coulomb branch deformations, $\chi(r)$ and
$\al(r)$, of the $\cN=4$ super-Yang-Mills theory along the RG flow:
\eqn\defofcrh{
c\eql\cosh(2\chi)\,,\qquad 
\rho\eql\exp(\al).}
They satisfy a system of first-order differential (flow) equations:
\eqn\flowcrho{
{dc\over dr}\eql \rho^4\,(1-c^2)\,,\qquad {d\rho\over dr}\eql
{1\over 3}\,\Big({1\over \rho}-c\,\rho^5\Big)\,.
}
whose general solution  is given by \refs{\PilchUE,\BrandhuberCT}:
\eqn\gensolfl{
\rh^6\eql c ~+~ (c^2-1)\,
\left[\,\gamma+{1\over 2}\log\left({c-1\over c+1}\right)\,\right]\,,
} 
where $\gamma$ is a constant of integration that parametrizes
different flows.

In the solution of the IIB supergravity,  all the bosonic
fields, the metric, $g_{MN}$, the dilaton/axion, $\tau$, the five-index 
tensor, $F_{(5)}$, and the three index tensor, $G_{(3)}$, are
non-vanishing. The metric is diagonal:
\eqn\ntwometric{
ds^2\eql \Om^2\,(dx_\mu dx^\mu)-\left(V_1^2 dr^2+V_2^2 d\th^2+
V_3^2(\si^1)^2+V_4^2 ((\si^2)^2+(\si^3)^2)+V_5^2 d\phi^2\right)\,,
}
with the functions
$\Om(r,\th)$ and $V_a(r,\th)$, $a=1,\ldots,5$,  given by:
\eqn\formofa{
\Om(r,\th) \eql 
{\ccr^{1/8} \rhr^{3/2} \XXone^{1/8} \XXtwo^{1/8}\over 
(\ccr^2-1)^{1/2}}\,,
}
and
\eqn\formofvs{\eqalign{
V_1(r,\th) 
&\eql {\ccr^{1/8} \XXone^{1/8}\XXtwo^{1/8}\over \rhr^{1/2}}\,,\cr
V_2(r,\th)&\eql {\XXone^{1/8}\XXtwo^{1/8}\over
c^{3/8}\rh^{3/2}}\,,\cr
V_3(r,\th)&\eql {\rhr^{3/2}\XXone^{1/8}\over 
\ccr^{3/8} \XXtwo^{3/8}}\,\cos\th \,,\cr
V_4(r,\th)&\eql {\ccr^{1/8}\rhr^{3/2}\XXtwo^{1/8}\over \XXone^{3/8}}\,\cos\th
\,,\cr
V_5(r,\th)&\eql {\ccr^{1/8}\XXone^{1/8}\over \rhr^{3/2} \XXtwo^{3/8}}\,
\sin\th\,,\cr
}}
where 
\eqn\defofX{
X_1(r,\th)\eql \cos^2\th+\ccr\,\rhr^6\sin^2\th\,,\qquad X_2(r,\th)\eql
\ccr\, \cos^2\th+\rhr^6\sin^2\th\,. }
The $SU(2)$ Maurer-Cartan forms $\si^i$, $i=1,2,3$, are normalized by
$d\si^1=2\, \si^2\wedge \si^3$. (Note that this normalization differs
by a factor of two from that used in \GNW.)  The metric has the Poincar\'e 
invariance along the brane directions, $x^\mu$, and is also manifestly
$SU(2)\times U(1)^2$ invariant, where the first $U(1)$ rotates $\si^2$
and $\si^3$, while the second $U(1)$ is a phase rotation in the angle
$\phi$. 

We alert the reader that the form of the metric \ntwometric\ differs
slightly from the one in \PilchUE\ in that we have combined all the
warp factors in the metric along the brane into a single function,
$\Om(r,\th)$. The orthonormal frames, $e^M$, $M=1,\ldots,10$, are the
same as in \PilchUE.

The dilaton/axion fields $(\Phi,C_{(0)})$ form a complex scalar, which
is related to the supergravity field, $B$, in the $SU(1,1)$ basis by
\eqn\dilsltwo{
\tau ~\equiv~ C_{(0)}+i\, e^{-\Phi}\eql i\left({1-B\over 1+B}\right)\,.
}
The latter is explicitly given by
\eqn\solforb{
B(r,\th,\phi)
\eql\left({b^{1/4}-b^{-1/4}\over b^{1/4}+b^{-1/4}}\right)\, e^{2i\phi}\,,
\qquad
b(r,\th)\eql {\ccr\,\XXone\over \XXtwo}\,.}
We also recall that the scalar one-form, $P$, and the connection form,
$Q$, are defined by:
\eqn\pandq{
P\eql {dB\over 1-BB^*}\,,\qquad Q\eql {1\over 2i}\,{B\,dB^*-B^*\,dB\over 
1-BB^*}\,.
}

The RR four-form potential, $C_{(4)}$, and the corresponding 
five-index field strength, $F_{(5)}$, are:
\eqn\cfourandffive{
C_{(4)}\eql w \,dx^0\wedge dx^1\wedge dx^2\wedge dx^3\,,
\qquad F_{(5)}\eql dC_{(4)}+*dC_{(4)}\,,
}
where the function $w(r,\th)$ can be written in the form
\EvansCT \foot{We correct here the misprint in \PilchUE, brought to our 
attention by the authors of \BuchelCN\ and \EvansCT.}: 
\eqn\defofw{
w(r,\th)\eql {\Om^4\over 4}\, { X_1^{1/2}\over c\,X_2^{1/2}}\,.
}

Finally, the three-index tensor field, $G_{(3)}$, is:
\eqn\theoldg{
G_{(3)}\eql (1-BB^*)^{-1/2}\,(dA_{(2)}-B\, dA_{(2)}^*)\,.
}
with the  two-form potential, $A_{(2)}$,  given by:
\eqn\theaaone{
A_{(2)}\eql e^{i\phi}\,\left(
 a_1\,d\theta\wedge \si^1+a_2\,\si^2\wedge \si^3+
a_3\,\si^1\wedge d\ph\right)\,,
}
where
\eqn\theas{\eqalign{
a_1(r,\th)&\eql -{i\over c}\, { (c^2-1)^{1/2}} \, \cos\th\,,\cr
a_2(r,\th)&\eql  i\,{\rh^6\over \XXone} \, (c^2-1)^{1/2}
\,\cos^2\th\,\sin\th\,,\cr
a_3(r,\th)&\eql -{1\over \XXtwo}\,  (c^2-1)^{1/2}
\,\cos^2\th\,\sin\th\,.\cr
}}
We note that $G_{(3)}$ has only six non-vanishing components when
expanded in the orthonormal basis, namely
\eqn\nonvangs{
G_{57\,10}\,,\qquad G_{67\,10}\,,\qquad G_{89\,10}\,,
\qquad
G_{567}\,,\qquad G_{589}\,,\qquad G_{689}\,, }
where, for $\phi=0$, the first three are real while the remaining three
are purely imaginary.

\newsec{Supersymmetry}
\seclab\Supersyymetry

The supersymmetry variations for  the gravitino, $\psi_M$, and 
the spin-${1 \over 2}$ field, $\lambda$, in IIB supergravity read \SchwarzQR:
\eqn\susytrpsi{
\delta\psi_M\eql D_M\ep+{i\over 480}F_{PQRST}\,\ga^{PQRST}\ga_M\,\ep
+{1\over 96}\left(\ga_M{}^{PQR}-9\,\delta_M{}^P\ga^{QR}\right)\,G_{PQR}
\,\ep^*\,,
}
and
\eqn\susytrla{
\delta\lambda\eql i\,P_M\,\ga^M\epsilon^*-{i\over 24} \,G_{MNP}\,
\ga^{MNP}\,\ep\,,
}
where $\ep$ is a complex chiral spinor satisfying%
\foot{We use the same notation and $\gamma$-matrix conventions as in
\SchwarzQR. Also, see appendix A.}
\eqn\chirsp{
\ga^{11}\ep\eql-\ep\,.
} 

The conditions for unbroken supersymmetry are
$\delta\psi_M=\delta\lambda=0$, which gives a combination of algebraic
and first order differential equations for the Killing spinor,
$\epsilon$.   We write these equations schematically as 
\eqn\killspeqs{
\partial_M\ep\eql\Delta_M\,\ep\qquad \hbox{and}\qquad 
\Delta_{1/2}\,\ep\eql0\,.
} 
Here $\partial_M$ denotes the partial derivative $\partial/\partial
x^M$, except for the $SU(2)$ directions, where we take $\partial_M$,
$M=7,8,9$, to be the $SU(2)$ invariant (Killing) vector fields dual to the
Maurer-Cartan forms, $\si^i$, $i=1,2,3$, respectively. The operators
$\Delta_M$ and $\Delta_{1/2}$ are purely algebraic and depend on all
background fields. In particular, the dependence on the metric and the
dilaton/axion in \susytrpsi\ arises from the connection terms in the
covariant derivative:
\eqn\covder{
D_M\ep\eql \partial_M\ep+{1\over 4}\om_{MPQ}\ga^{PQ}\ep-{i\over 2}
Q_M\ep\,.
}

The operators $\Delta_M$ and $\Delta_{1/2}$ will in general involve
the operation of complex conjugation, which will be denoted by
`$*$'.    In practice it is often convenient to pass to a real
realization of operators by decomposing spinors into the real and
imaginary parts.

It has been emphasised recently  (see, for example,
\refs{\GauntlettSC\GauntlettNW{--}\GauntlettFZ}) that a lot of information about
a supersymmetric background can be recovered from the Killing spinors
by constructing canonical vector fields and differential forms
associated with them -- the so-called G-structures. Here, we will
concentrate on vector fields. That is, given two Killing spinors, $\ep^a$ and 
$\ep^b$,  consider:
\eqn\killvect{
K_{ab}^M\eql \overline{\ep^a} \ga^M\ep^b \,, } 
which, as expected, is a  Killing vector of the metric.  More generally,
one can consider the differential forms:
\eqn\Forms{\Omega^{(ab)}_{M_1M_2\dots M_p} ~\equiv~\overline{\ep^a} 
\, \gamma_{M_1M_2\dots M_p}\, \ep^b \,.}
Such forms can be used to
partially determine the potentials for the antisymmetric tensor fields \GNW. 

At this point the standard procedure would be to examine integrability
conditions for the equations,  however, since we are interested in
an explicit form of the Killing spinors, we will proceed directly with
the solution by first recalling the standard calculation at the
maximally supersymmetric point (see, for example,
\refs{\LuTH,\LuNU}).  This will set a proper stage for the
discussion of the general case.

\subsec{Supersymmetry at the $\cN=8$ point}

The maximally supersymmetric point is the $AdS_5\times S^5$ solution
of the IIB supergravity, which in the present set-up is recovered by
taking the limit:
\eqn\adslimit{
c(r)~\longrightarrow~1\,,\qquad
\rh(r)~\longrightarrow~1\,,\qquad
\Om(r,\th)~\longrightarrow~ e^r\,.
}
The only background fields are the metric and the five-index tensor given by:
\eqn\adsmetric{
ds^2\eql e^{2r}(dx_\mu dx^\mu)-dr^2-\left(d\theta^2+
\cos^2\th\,((\si^1)^2+\ldots+(\si^3)^2)+\sin^2\th\,d\phi^2\right)\,,
}
\eqn\adsfive{
F_{(5)} \eql e^{4r}\,dx^0\wedge dx^1\wedge dx^2\wedge dx^3\wedge dr+
\sin\th\,\cos^3\th \, d\th\wedge \si^1\wedge\si^2\wedge\si^3\wedge d\phi\,.
}

As in the standard calculation we  write $\De_M$
as products of projectors and invertible operators using \chirsp\
along the way, but no other conditions. We obtain:
\eqn\delone{\eqalign{
\De_1 &\eql -{1\over 2}\,e^{r}\,\ga^1  \ga^5\,(1+i\,\ga^1\ldots\ga^4)\,, \cr
\De_j &\eql {1\over 2}\,e^{r}\,\ga^j \ga^5\,(1+i\,\ga^1\ldots\ga^4)\,, \quad j=2,3,4 \cr
}
}
and
\eqn\delfive{
\De_5\eql -{i\over 2}\,\ga^1\ldots\ga^4\,,
}
\eqn\delth{
\De_6\eql {i\over 2}\,\ga^1\ldots\ga^4\ga^5\ga^6\,.
}
For the $SU(2)$ directions we have:
\eqn\delsutwo{
\De_i\eql {1\over 2}\,\ga^j\ga^k\,({\cI}+i\,\cos\th \,\ga^6\ga^{10}+\sin\th 
\ga^1\ldots\ga^4\ga^5\ga^{10})\qquad i,j,k=7,8,9\,,
}
and, finally, 
\eqn\Deten{
\De_{10}\eql {i\over 2}\,(i\,\cos\th\,\ga^6\ga^{10}+ \sin\th\,
\ga^1\ldots\ga^4\ga^5\ga^{10})\,.
}
While this point has sixteen supersymmetries (in the presence of the brane), 
we are interested in isolating the eight supersymmetries that will remain unbroken 
under the flow solution described in the previous section.  These eight supersymmetries 
transform as a doublet under the residual $SU(2)$ $\cR$-symmetry.  Thus, 
we are interested in determining supersymmetries that are invariant 
along the brane, \ie \ $\ep$ is independent of $x^\mu$, and 
transform non-trivially under $SU(2)$.   Using \delone\ and \delsutwo\ 
these two conditions are equivalent to  two algebraic equations:
\eqn\projeqs{
\Pi_0^{(1)}\, \ep\eql \ep\,,\qquad \Pi^{(2)}(\th)\, \ep\eql\ep\,,
}
where
\eqn\projone{
\Pi_0^{(1)}\eql {1\over 2}\,({\cI}-i\ga^1\ga^2\ga^3\ga^4)\,.
}
and 
\eqn\prtwop{
\Pi^{(2)}(\th)\eql {1\over 2}\left({\cI}+
i\,(\sin\th\,\ga^5+\cos\th\,\ga^6)\,\ga^{10}\right)\,.
}
These are mutually commuting projectors and thus define a space
of eight (real) supersymmetry parameters. Note that $\Pi_0^{(1)}$ is
the standard projector parallel to the $D3$-brane.  

Using 
\eqn\rotid{
\cos\th\,\ga^6+\sin\th\,\ga^5\eql \cO(\th)\,\ga^6\,\cO(\th)^{-1}\,,
}
where
\eqn\theomat{
\cO(\th)\eql \cos{\th\over 2}-\sin{\th\over2 }\,\ga^5\ga^6\,,
}
we also have
\eqn\pitworot{
\Pi^{(2)}(\th)\eql \cO(\th)\,\Pi_0^{(2)}\,\cO(\th)^{-1}\,,\qquad
\Pi_0^{(2)}\eql {1\over 2}(\cI+i\ga^6\ga^{10})\,.
}
Given that $\cO(\th)$ satisfies the equation
\eqn\eqforo{
\partial_\th\,\cO(\th)\eql -{1\over 2}\,\ga^5\ga^6\,\cO(\th)\,.
}
it is now straightforward to integrate all the equations with the result
\eqn\epsres{
\epsilon\eql e^{r/2}\,e^{i\ph/2}\,\cO(\th)\,\epsilon_0\,,
}
where $\epsilon_0$ depends only on the $SU(2)$ directions and satisfies:
\eqn\projoneps{
\Pi_0\,\ep_0\eql\ep_0\,,\qquad \Pi_0\equiv \Pi_0^{(1)}\Pi_0^{(2)}\,.
}
If one combines \delsutwo\ with \prtwop\  one sees that the
action of $SU(2)$ is generated by the  products of gamma matrices 
$\gamma^{67}$, $\gamma^{68}$ and
$\gamma^{78}$.  However, because of the projection matrices, there are several 
ways to express this action on $\ep_0$.  Indeed, the $SU(2)$ action
can be generated by the matrices:
\eqn\sutwogens{
t_1\eql \ga^8\ga^9\,,\qquad t_2\eql -\ga^5 \ga^8\,,\qquad 
t_3\eql -\ga^5\ga^9\,,
}
which can be related to the more canonical set using the identities:
\eqn\idforfive{
\ga^7\ga^9\,\Pi_0\eql \ga^5\ga^8\, \Pi_0\,,
\qquad
\ga^7\ga^8\,\Pi_0\eql -\ga^5\ga^9\, \Pi_0\,.
}
We will find the generators \sutwogens\ more convenient in the 
$\cN=2$ flow solution.  The dependence of the Killing spinors
on the Euler angles,  $\al^i$, of the $SU(2)$ can thus be obtained
by exponentiating these matrices in much the same manner as was done
in \refs{\PopeJP,\GNW}.

Finally,  consider the vector field:
\eqn\killvectfield{
K\eql \overline\ep\ga^M\ep\,\partial_M \eql
e^r\,\overline\ep_0\cO(\th)^\dagger\ga^M\cO(\th)\ep_0\,\partial_M\,.}
Using \projeqs,  we find that for $M\geq 5$ we have:
\eqn\smallcalc{\eqalign{
K_M&~\propto~ \overline\ep_0\ga^M\ep_0\cr
&\eql \overline\ep_0(1-\Pi_0^{(1)})\,\ga^M\,\Pi_0^{(1)}\ep_0\cr
&\eql 0\,.\cr}}
Thus, the non-vanishing (frame) components are:
\eqn\component{
(K^M)\eql e^{r} \,k^M\,,\qquad M=1,\ldots,4\,, } 
where $k^M$ are constants as the dependence on $\theta$, $\phi$ and 
the Euler angles drops out in \killvectfield. 
We thus see explicitly that  the Killing spinors yield only those
Killing vectors which correspond to the ``trivial'' symmetry of the
metric \adsmetric, namely translations along the brane. However, this
result is quite non-trivial if we reverse the order of reasoning and
observe that the $r$-dependence of the Killing spinors is completely
fixed once we require that $K$ yields some non-zero Killing vectors of
the metric. In fact, this observation will be crucial for the
calculation of the Killing spinors in the next section.

\subsec{Supersymmetry along the $\cN=2$ flow}

We now return to the  $\cN=2$ background in Section
\rgflowsolution. Our general strategy will be exactly as above, and will
be parallel to corresponding approach in $M$-theory \GNW. 
First we consider those equations in \killspeqs\ that are purely
algebraic and construct the projector onto the space of
solutions. Then we use the relation with the Killing vectors to
integrate the remaining differential equations explicitly.

The resulting Killing spinors must continously reduce to the ones we
have already found at the $\cN=8$ point.   Because of the Poincar\'e 
symmetry along the brane, and because of the residual
$\cR$-symmetry we know that: (i) the Killing spinors, $\ep$,
are constant along the brane,  (ii) the dependence on 
rotational coordinates, $x^7,x^8,x^9$, can be obtained by exponentiating
suitably defined $SU(2)$ generators, and (iii) the $\phi$ dependence of the
Killing spinors appears through the simple phase as in \epsres.  
On a more mechanistic level, one can verify {\it a postiori}  that this
simplyfying Ansatz yields all unbroken supersymmetries. One can also
easily check that this $\phi$-dependence, given the
phases in \solforb, \pandq, \theaaone\ and  \theoldg, is in fact the only that is 
consistent with the appearance of $\ep$ and $\ep^*$ in  
the spin-${1 \over 2}$ variation, \susytrla. 

Our Ansatz yields  six algebraic equations with  four of them 
manifestly equivalent:
\eqn\thesame{
\ga^1\De_1\,\ep\eql\ldots\eql\ga^4\De_4\,\ep ~=~ 0 \,.
} 
Guided by the same strategy as in  the construction of the lift in
\PilchUE, we now seek a linear combination of $\De_1$, $\De_{10}$ and
$\De_{1/2}$ in which the dependence on the three-index tensor field
cancels out. We find that such a combination is indeed possible and
that the result is quite simple, namely:
\eqn\nicecombination{
2 \big(\,\ga^1\De_1+\ga^{10}(\De_{10}-{i\over 2})\big)\,\ep
-\big(\De_{1/2}\,\ep\big)^*\,
\eql 
\big(\cos\th\,\ga^6+\sqrt{c}\,\rh^3\sin\th\,\ga^5+i\,
\sqrt{X_1}\,\ga^{10}\big)\ep\,,
}
where the left-hand side is required to vanish. Here we have used
the background values for the fields, but it is easy to see that only
the metric and the dilaton/axion contribute to the right hand side.
We can now rewrite the resulting equation in terms of a
projector. Indeed the vanishing of the right hand side of
\nicecombination\ is equivalent to:
\eqn\thepojtw{
\Pi^{(2)}(\al)\,\ep\eql\ep\,,
}
where the projector $\Pi^{(2)}(\al)$ is the same as in \prtwop\ 
with the angle $\al(r,\th)$ determined by
\eqn\rotang{
\tan\al\eql \sqrt{c}\,\rh^3\,\tan\th\,,
} 
or, equivalently,
\eqn\cosinal{
\cos\al\eql {\cos\th\over X_1^{1/2}}\,,\qquad
\sin\al\eql {\sqrt{c}\,\rh^3\,\sin\th\over X_1^{1/2}}\,.
}

Now, rather than chasing other linear combinations to obtain further
projectors, we proceed quite directly. We can simplify the spin-${1 \over 2}$
variation by restricting it to the subspace of spinors satisfying
\thepojtw.   After some algebra it is then possible to write down the 
general form of the solution and construct the corresponding projector:
\eqn\Projone{
{1\over 2}\left(\cI- i\,\ga^1\ga^2\ga^3\ga^4\,
(p_1(r,\th)+p_2(r,\th)\ga^7\ga^{10}*\,)\right) \,,
}
where 
\eqn\theps{
p_1(r,\th)\eql {X_1^{1/2}\over c^{1/2}X_2^{1/2}}\,,\qquad
p_2(r,\th)\eql {(c^2-1)^{1/2}\,\cos\th\over c^{1/2}X_2^{1/2}}\,.
}
Introduce an operator
\eqn\calost{
\cO^*(\be)\eql \cos{\be\over 2}+\sin{\be\over 2}\,\ga^7\ga^{10}\,*\,.
}
Then \Projone\ can be simply written as:
\eqn\rotpione{
\Pi^{(1)}(\be)\eql 
\cO^*(\be)\,\Pi_0^{(1)}\,\cO^*(\be)^{-1}\,,
}
where $\Pi_0^{(1)}$ is defined in \projone\ and the 
deformation angle $\be(r,\th)$ is determined by:
\eqn\thebetang{
\cos\be\eql {X_1^{1/2}\over c^{1/2}\,X_2^{1/2}}\,,
\qquad
\sin\be\eql- {(c^2-1)^{1/2}\,\cos\th\over c^{1/2}\,X_2^{1/2}}\,.
}
It is now easy to verify that  {\it all} algebraic equations have been 
solved, and thus  $\ep$ must satisfy:
\eqn\fullproj{
\Pi(\al,\be)\,\ep\eql \ep\,,\qquad
\Pi(\al,\be)~\equiv~\Pi^{(1)}(\be)\,\Pi^{(2)}(\al)\,.
}

It is interesting that in fact \fullproj\ follows from the spin-${1 \over 2}$
variation alone.  We have verified this by brute force algebra after
we have obtained the projection condition using the simplifications
outlined above. It would be useful to have a more direct proof of this
observation.

The next step is to integrate explicitly the remaining first
order equations. We start with the $SU(2)$ directions and first verify that:
\eqn\presutwo{
\De_i\,\Pi(\al,\be)\eql \Pi(\al,\be)\, \De_i\,\Pi(\al,\be)\,,
}
which shows, as expected, that the solutions to the algebraic
equations form a representation of $SU(2)$.  More specifically, we find
the identities:
\eqn\onsutwo{\eqalign{
\De_i\,\Pi(\al,\be)& \eql
\cO(\al)\,t_i\cO(\al)^{-1}\,\Pi(\al,\be)\cr
&\eql\Pi(\al,\be) \,\cO(\al)\,t_i\cO(\al)^{-1}\,,\cr} } 
where the generators $t_i$ are  given by \sutwogens.   This provides an
explicit action of $SU(2)$ on the solutions of \fullproj.  This can then
be exponentiated to yield the dependence of the solution upon the Euler angles.
Equivalently, this equation shows how to map the  $SU(2)$ dependence
at the maximally supersymmetric point onto the $SU(2)$ dependence anywhere
along the flow.

Consider spinors of the form $\Pi(\al,\be)\cO(\al)\ep_0$, where
$\ep_0$ satisfies \projoneps\ and transforms under $SU(2)$ with the
generators $t_i$. By construction, such spinors solve 
\killspeqs, except for the two equations along the $r$ and $\th$ directions.
Since we expect to find eight (real) independent components of the
Killing spinors corresponding to the unbroken $\cN=2$ supersymmetry,
and since the dimension of the range of the projector $\Pi(\al,\be)$
is equal to eight, it remains only to fix the overall normalization of
the solution.  Rather than trying to integrate the remaining equations
explicitly, we take a shortcut and require that using \killvect\ the
solution gives rise to at least one Killing vector along the
brane. This leads to an prescription:
\eqn\finres{\eqalign{
\ep \eql {\Om^{1/2}e^{i\phi/2}\over \cos(\be/ 2)} 
\Pi(\al,\be)\,\cO(\al)\ep_0
&\eql  {\Om^{1/2}e^{i\phi/2}\over \cos(\be/ 2)} 
\cO(\al)\cO^*(\be)\Pi_0^{(1)}
\cO^*(\be)^{-1}\,\ep_0\,,\cr}
}
with $\ep_0$ satisfying
\eqn\projonepzero{
\Pi_0^{(1)}\Pi_0^{(2)}\,\ep_0\eql \ep_0\,.
} 
It is then quite straightforward to check that \finres\ is
indeed the general solution to the Killing spinor equations
\killspeqs\ for this IIB supergravity background.

\subsec{Comments: New coordinates and the ``Killing structure''}

We have now shown explicitly that our solution is indeed $\cN=2$
supersymmetric. By comparing \finres\ with \epsres\ we see that the
deformation of the Killing spinor as we go away from the maximally
supersymmetric point is encoded in the operator $\cO^*(\beta)$, which
rotates the projector $\Pi_0^{(1)}$ into $\Pi^{(1)}(\beta)$. The other rotation, 
$\cO(\al)$, is merely an
artifact of a particular choice of orthonormal frames for the metric
and clearly can be removed by rotating $e^5$ and $e^6$, the frames
along the $r$ and $\theta$ coordinates, to a new set of frames, $e^u$
and $e^v$, given by:
\eqn\newframes{
\left (\matrix { e^u\cr e^v\cr}\right) \eql
\left(\matrix{ \cos\al &  - \sin\al \cr 
	\sin\al & \cos\al \cr}\right)
\left (\matrix { e^5\cr e^6\cr}\right)\,.
} 
Obviously, one would like to know whether there is a corresponding
change of variables, $(r,\th)\rightarrow(u,v)$, which induces 
this rotation. 

A similar issue appears in \GNW, where it was shown that such a
change of variables does indeed exist. Using the parallels between the
solution in \GNW\ and the present one, we find that the new coordinates are
given by:
\eqn\defuv{
u(r,\theta)\eql {\rhr^3\,\cos\theta\over (c^2-1)^{1/2}}\,,\qquad
v(r,\theta)\eql {\sin\th\over (c^2-1)^{1/2}}\,.  } 
In the next section we will show that the existence of these
coordinates is a direct consequence of the supersymmetry of the
background. 

In terms of those new coordinates the metric, and other background
fields, have a relatively simple structure. In particular, the metric becomes
\eqn\newmetr{\eqalign{
ds^2\eql  & \Om^2\,\left(dx_\mu dx^\mu\right)~-~
{\Om^{-2}\over \cos\be }\,\left(du^2+{1\over c} 
dv^2 + u^2((\si^2)^2+(\si^3)^2)\right) \cr & -
\Om^{-2}\,\cos\be\, \,\left(u^2(\si^1)^2+cv^2 (d\phi)^2\right)\,,\cr}
}
where $\Om(u,v)$, $c(u,v)$ and $\be(u,v)$ are now  functions of
$u$ and $v$. We will label the orthonormal frames for this metric as 
before with the correspondence  $e^u\leftrightarrow e^5$ and 
$e^v\leftrightarrow e^6$.  Also note that in these new coordinates:
\eqn\omsimp{\Omega^4 ~=~ u^2 {\cos \beta \over \sin^2 \beta} \,.}

Finally, we find that \killvect\ gives rise to five Killing vectors,
which, with respect to the orthonormal frame above, have the 
components:
\eqn\killcomp{
(K^M)\eql \Omega\, (k_1,\, k_2,\,  k_3,\,  k_4,\, 0,
	\, 0,\, k_7\,\sin \be,\,  0,\, 0,\, 0)\,,
}
where $k_i$ are constants. From \newmetr\ and \omsimp\ one
can see that the coordinate components, $K^\mu$, are all constants.
Thus, as we move away from the $\cN=8$
point there is one additional Killing vector, which corresponds to the
$U(1)$ symmetry that rotates $\si^2$ and $\si^3$.  The fact that this
new Killing vector is not forbidden by \smallcalc\ is a direct consequence 
of the deformation of $\Pi_0^{(1)}$ in \projone\ to  $\Pi^{(1)}$ in \Projone\
via the rotation \rotpione.  

It is now an interesting problem to find to what extent the metric or the
entire background is determined by the form of the Killing spinors
\finres. We will study this more general problem in the 
next section.

\newsec{Generalized $\cN=2$ backgrounds}

In this section we discuss in more detail the constraints on the
background imposed by the requirement of $\cN=2$ supersymmetry
({\it i.e.} eight supersymmetries). More
specifically, we want to detrmine the most general $\cN=2$
supersymmetric solution of the chiral IIB supergravity arsing from the
following Ansatz:
\smallskip
\item{(i)} The metric is diagonal of the form \ntwometric, where the 
$\Om(r,\th)$ and $V_a(r,\th)$, $a=1,\ldots,5$ are arbitrary functions.
\item{(ii)} The dilaton/axion is given by \dilsltwo\ and \solforb\ 
with an arbitrary function $b(r,\th)$.
\item{(iii)} The five-index tensor has only four non-vanishing components:
$F_{12345}$, $F_{12346}$, and the ones related by the
condition of self-dualty.  These components  are to be functions of $r$ and 
$\theta$ alone.
\item{(iv)} The non-vanishing components of the three-index tensor are,
at most, those listed in \nonvangs. Their dependence on $\phi$ is only
through the overall phase, $\exp(i\phi)$, and they satisfy the same
reality conditions at $\phi=0$ as those in \nonvangs.
\item{(v)} The Killing spinors of the unbroken supersymmetry have the form 
\finres, where $\al(r,\th)$ and $\be(r,\th)$ are arbitrary functions.
\item{(vi)} The Killing spinors transform under $SU(2)$ with the  
same action as in \onsutwo.
\item{(vii)} The Killing vectors that are generated by 
bilinears of Killing spinors include the $U(1)$ rotation as in  \killcomp.

\subsec{Solving the Ansatz}

We will solve our problem in two steps. First we insert the general
Ansatz for the background fields and the Killing spinors into the
supersymmetry equations \killspeqs\ and show that all unknown
functions, $\Om(r,\th)$, $V_a(r,\th)$, $b(r,\th,\ph)$, $\beta(r,\th)$
and the components of the three-index and five-index tensors are completely
determined in terms of a single function, $c(r,\theta)$.  The dependence
on the angle, $\al(r,\th)$,  can be removed by a suitable change of
coordinates. The Bianchi identities for the tensor fields turn out to
be equivalent to a second order partial differential equation for c. In
the second step we verify that all field equations are satisfied.

We begin by performing a rotation by the angle $\al$ from
$(e^5,e^6)$ to $(e^u,e^v)$ as in \newframes.  Then
\eqn\newvecf{
E_u\eql {\cos\al\over V_1}\,{\partial\over \partial r} -
{\sin\al\over V_2}\,{\partial\over \partial \th}\,,
\qquad
E_v\eql {\sin\al\over V_1}\,{\partial\over \partial r} +
{\cos\al\over V_2}\,{\partial\over \partial \th}\,,
}
are vector fields dual to $e^u$ and $e^v$, respectively. We denote the
corresponding tensor indices in the new frame by $u$ and $v$.

We can now rewrite the supersymmetry equations \killspeqs\ in the new
basis and in terms of derivatives $E_u$ and $E_v$. In Appendix C we
have compiled a list of independent equations and discuss briefly
how they arise.

We note that the assumption (vii) about the existence of the 
$U(1)$ Killing vector of the form \killcomp\ is equivalent to 
the following condition on the functions $V_3$ and $V_4$:
\eqn\ratiothfo{
{V_3\over V_4}\eql \cos(\be)\,.
}
Then, using \eqca, we obtain
\eqn\thevth{
V_4\eql {\sin(2\be)\over \cos(\be)\,G_{89\,10}-i\,G_{567}}\,.
}

Guided by the form of the metric \newmetr,  we define a function
\eqn\defofu{
u(r,\th)\eql \Om\,V_4\,(\cos\be)^{1/2}\,.
}
Using  \eqba, \eqda, \eqdd\ and then \thevth, one can check that
\eqn\eqsforu{
E_u\,u\eql \Om\,(\cos\be)^{1/2}\,,\qquad
E_v\,u\eql 0\,.
}
Similarly,  we define
\eqn\defofv{
v(r,\th)\eql {\Om\,V_5\over c^{1/2}\,(\cos\be)^{1/2}}\,,
}
where 
\eqn\defofc{
c(r,\th)\eql {1\over\cos\be}\,\left({1+B\over1-B}\right)_{\phi=0}\,.
}
Using \eqba, \eqda\ and \eqdb, we find
\eqn\eqsforv{
E_u\,v\eql 0\,,\qquad E_v\,v\eql \Om\,c^{1/2}\,(\cos\be)^{1/2}\,.  }
This shows that upon the change of coordinates from $(r,\th)$ to
$(u,v)$ given in \defofu\ and \defofv, the forms $e^u$ and $e^v$ are
proportional to $du$ and $dv$:
\eqn\euev{
e^u\eql {\Om^{-1}\over (\cos\be)^{1/2}}\,du\,,\qquad e^v\eql
{\Om^{-1}\over c^{1/2}\,(\cos\be)^{1/2}}\,dv\,, } 
and thus the metric has a diagonal form in the new coordinates as well. 

Thus, just as in the special solution in Section 3,
the dependence on the angle, $\alpha$, can be removed by a change of
coordinates, and the existence of the new coordinates, $u$ and
$v$, which essentially is the content of the equations \eqsforu\ and
\eqsforv, is a consequence of supersymmetry.  From now on we will 
work in the new coordinates and define $V_u$ and $V_v$ as the
coefficients of $du$ and $dv$ in \euev, respectively. It also follows
from \ratiothfo-\euev\ that in the new coordinates the metric has
precisely the form \newmetr, where $\Om(u,v)$, $\beta(u,v)$ and
$c(u,v)$ are, at this point, arbitrary functions, which we will now
restrict further by solving the equations (a)-(f) in Appendix C.

First, we  evaluate 
\eqn\whatwea{
E_uc\eql \Om\,(\cos\be)^{1/2}\,{\partial_u c}\,,}
and
\eqn\whatweb{ 
E_v c\eql \Om\,c^{1/2}\,(\cos\be)^{1/2}\,{\partial_v c}\,.
} 
Using \defofc, \eqaa-\theqs\ and \eqba, we find
\eqn\deroncu{
E_uc\eql -i\,c\,\tan(\be)\,G_{567}\,,
}
and 
\eqn\deroncb{
E_vc\eql {\Om\over v\,c^{1/2}\,(\cos\be)^{3/2}}\,(1-c^2\,\cos^2\be)\,, }
where, in the last equation, all dependence on the the antisymmetric 
tensor field cancelled
out and we used \defofv\ to eliminate $V_5$. Combining \whatweb\ and 
\deroncb, we can now determine $\be$ in terms of $c$:
\eqn\solforbe{
\be(u,v)\eql -\arctan\left(c^2+v\,c\,\partial_vc-1\right)^{1/2}\,.
}

Next we use \thevth, \defofu, \deroncu\ and \eqba\ to express $G_{567}$,
$G_{u89}$, $G_{v89}$ and $G_{89\,10}$ in terms of $\Om$ and $c$. Then
we substitute  the result in \eqda\ and solve for 
$\partial_u\Om$ and $\partial_v\Om$ in terms of $u$, $v$, $c(u,v)$ and
its partial derivatives.  We verify that the system is integrable and obtain:
\eqn\solofom{
\Om(u,v)\eql {u^{1/2}\,c^{1/8}\,(c+v\,\partial_vc)^{1/8}\over
(c^2+v\,c\,\partial_vc-1)^{1/4}}\,,
}
together with 
\eqn\solforB{
B(u,v,\phi)\eql \left( 
{c-c^{1/2}\,(c+v\,\partial_vc)^{1/2}\over
c+c^{1/2}\,(c+v\,\partial_vc)^{1/2}}
\right)\,e^{2i\phi}\,,
}
which follows from \defofc\ and \solforbe.  This
shows that the metric and the dilaton/axion are given in terms of a
single function $c(u,v)$. We also verify that all equations (a)-(f) in
Appendix C are now satisfied for an arbitrary $c(u,v)$.

Further, using \eqee, \eqac\ and the relations obtained above, we can
express all the components of the antisymmetric tensors field
strengths, $G_{(3)}$ and $F_{(5)}$, in terms of $c(u,v)$ and its
derivatives. In order to compute the corresponding potentials,
$C_{(4)}$ and $A_{(2)}$, we first check whether Bianchi identities  
\SchwarzQR\
\eqn\bianffive{
dF_{(5)}+{i\over 8}\,G_{(3)}\wedge G_{(3)}^*\eql0\,,
}
and
\eqn\baingth{ d\left[\, (1-BB^*)^{-1/2}\,(G_{(3)}+B\,
G^*_{(3)}) \,\right]\eql0 \,,
}
are satisfied. 

The result of the calculation can be succinctly expressed in terms of
\eqn\thelapl{
\cL(c)\eql  {\partial\over\partial u}\,\left({v^3\over u}\,
{\partial c\over\partial u}\right) ~+~
{\partial\over\partial v}\,\left({v^3\over u}\,c\,
{\partial c\over\partial v}\right)\,.
}
The equations resulting from the two Bianchi identities read
\eqn\bianone{
v\,c\,\partial_v\cL(c)+
(v\partial_v c-c)\,\cL(c)\eql 0\,,
}
and
\eqn\biantwo{
\cZ(c)\,\cL(c) \eql 0\,,\qquad 
\cZ(c)\,(v\partial_v\cL(c)-2\cL(c))\eql 0\,,
}
respectively, where
\eqn\defofz{
\cZ(c)\eql  2\,c^{1/2}\,(c+v\partial_v c)+(2c+c\partial_vc)(c+v\partial_v c)^{1/2} \,.
}
An obvious solution to those equations is obtained by setting
\eqn\theeqs{
\cL(c)\eql 0\,,
}
and so we will henceforth impose this on $c(u,v)$.

The explicit form of  $C_{(4)}$  is now simply obtained by setting,
\eqn\solofw{
w(u,v)\eql {1\over 4}\,\Om^4\,\cos\be\,,
}
where $w(u,v)$ is defined in  \cfourandffive\ and \defofw.
Finally, to calculate the two-form potential, $A_{(2)}$, we consider an
Ansatz of the form:
\eqn\newatens{
A_{(2)}\eql e^{i\phi}\,\left( 
a_1\,dv \wedge \si^1+a_2\,\si^2\wedge \si^3+
a_3\,\si^1\wedge d\ph \right)\,,
}
with arbitarary functions $a_i(u,v)$, $i=1,2,3$.
We find that, provided \theeqs\ is satisfied, $A_{(2)}$ is the desired 
potential if we set 
\eqn\theaaa{\eqalign{
a_1(u,v) &\eql  {i\over c} \,,\cr
a_2(u,v) &\eql  i\,{v\over v\,\partial_vc +c}\,,\cr
a_3(u,v) &\eql   -u\,v\,\partial_uc+2\,v\,c\,.\cr
}}

We have thus shown that all fields in our Ansatz are determined by the
 supersymmetry and Bianchi identities. We have also verified that all
 field equations are now satsified without any further conditions on
 $c(u,v)$. 

\subsec{Summary}

Here we pull together the key elements of the foregoing derivation
to show how a complete solution can be obtained trivially once one
finds a solution of the  ``master equation,'' defined by \theeqs\  and
 \thelapl.   

The metric functions, $V_u$, $V_v$ and $V_j$, $j=3,4,5$ can be read
off from \euev, \defofu, \defofv\ and \ratiothfo.  The resulting metric can then be
written in the form:
\eqn\metricfnl{\eqalign{
ds^2\eql \Om^2\,(dx_\mu dx^\mu)~-~ & \Om^{-2}\,\Big[H_1 \, \big(du^2 +  
u^2\,((\si^2)^2+(\si ^3)^2) \big) ~+~  H_1^{-1} \,u^2\,  (\si^1)^2 \cr & ~+~ 
 H_2  \, dv^2 ~+~ H_2^{-1} \, v^2\, d\phi^2\Big]\,,
}}
where
\eqn\Hdefs{H_1(u,v) ~\equiv~ {1 \over \cos \beta}  \,, \qquad H_2 (u,v) ~\equiv~ 
{1 \over c \, \cos \beta} \,.}
More importantly, from \Hdefs\ and \deroncb, we have:
\eqn\Hdefs{H_1 \, H_2^{-1} ~=~ c    \,, \qquad H_1\, H_2 ~=~ 
\partial_v ( v \,c)   \,,}
which shows that the $H_1$ and $H_2$ can be trivially generated from $c$.
The expression, \solofom, for  $\Omega$ may be written algebraically
in terms of $H_1$:
\eqn\omform{\Omega  ~=~ {u^{1/2}   \over   \big( H_1  ~-~ 
H_1^{-1}\big)^{1 / 4}}   \,.}
Similarly, the expression for the dilaton and axion may be re-written in
terms of $H_2$:
\eqn\dilaxform{B ~=~ {(1 ~-~ H_2)  \over  (1 ~+~ H_2) } \,e^{2 i \phi} \quad 
\Leftrightarrow \quad \tau ~=~  - { (H_2 \, \sin \phi  ~+~ i \, \cos \phi)\over  
(H_2 \, \cos \phi  ~-~ i \, \sin \phi)   }  \,.}
The expression, \solofw,  for the function that governs the $4$-form potential 
may be re-written as:
\eqn\resolofw{
w(u,v)\eql {1\over 4}\,\Om^4\,\cos\be ~=~{1\over 4}\,  {u^2   
\over   \big( H_1^2  ~-~ 
1\big) }   \,,}
while the expression for  the $2$-form potential is already 
given in terms of $c$ by \theaaa.  Note that:
\eqn\algas{\eqalign{a_1 ~=~ & i\, H_2 \, H_1^{-1}  \,,
 \qquad a_2 ~=~ i\, v\, H_1^{-1} \, 
H_2^{-1}  \,, \cr
a_3 ~=~  & - u^3 \, \partial_u \Big({v \over u^2}\, c\Big)\quad \Rightarrow
\quad  \partial_v a_3 ~=~- u^3 \, \partial_u 
\Big({1 \over u^2}\, H_1\, H_2\Big) \, \,.}}
The structure of this solution is manifestly very similar to that of \GNW.

\subsec{Some examples}

We have obtained a general class of $\cN=2$ solutions which are
determined by a single function $c(u,v)$ satisfying \theeqs. While the equation is
non-linear, one can easily find  some explicit solutions.
\medskip
\noindent
{\it Example 1.:\/} The original solution in Section~\rgflowsolution\
does indeed fall into this class of solutions.
We can use \defuv\ to obtain:
\eqn\rhosix{
\rh^6\eql {u^2\,(c^2-1)\over 1-v^2(c^2-1)}\,.
} 
Then we substitute the result in \gensolfl\ and differentiate with
respect to $u$ and $v$ to eliminate the integration constant, $\gamma$.
This yields a first-order system of equations for $c(u,v)$:
\eqn\fstordeqs{\eqalign{
{\partial c\over\partial u}&\eql 
{u\,(c^2-1)^2\,\left[v^2(c^2-1)-1\right]\over 
{(1 + v^2)^2+ u^2\,v^2\,c-2\,v^2\,( 1+v^2) \,c^2
	-2\,u^2\,v^2\,c^3+v^4\,c^4+u^2\,v^2\,c^5}
} \,,\cr
{\partial c\over\partial v}&\eql -{u^2\,v\,(c^2-1)^3\over
{(1 + v^2)^2+ u^2\,v^2\,c-2\,v^2\,( 1+v^2) \,c^2
	-2\,u^2\,v^2\,c^3+v^4\,c^4+u^2\,v^2\,c^5}
}
\,,\cr
}}
from which \theeqs\ follows directly.
\medskip

\noindent
{\it Example 2.:\/}  A simpler example can be generated by 
looking for solutions in which the two terms in \thelapl\ vanish
separately.   This yields a three-parameter family of solutions with
\eqn\threepsol{ c ~=~ \mu\,(1 + b\, u^2)\, \Big( 1 - {a \over v^2} \Big)^{1/2 } \,,}
for some constants $a$, $b$ and $\mu$.  One then finds 
\eqn\Hsol{ H_1 ~=~ \mu\,(1 + b\, u^2)  \,, \qquad H_2 ~=~ 
{v \over (v^2  - a  )^{1/2 }} \,.}
This is analogous to the solution found in \GNW.   For non-zero 
$a$ and $b$  the background is highly non-trivial.  Indeed, if
$b \ne 0$, the six-dimensional metric in the square brackets of
\metricfnl\ is asymptotic (at large distances) to an $S^1$ of constant radius 
non-trivially fibered over a flat $\IR^5$.   To have this six-metric asymptote
to a flat $\IR^6$ one must  $b=0$ and $\mu=1$, however this leads 
to singular expression for $\Om$.  One can arrive at a non-singular
result by taking $b=0$, $\mu = 1 + \eta$, and then taking the limit
$\eta \to 0$ while scaling $u, v$ and $x^\mu$ by suitable powers
of $\eta$.   The end result is a metric of the form:
\eqn\metricfnl{ 
ds^2\eql H_0^{-{1\over 2}} \,(dx_\mu dx^\mu)~-~ H_0^{{1\over 2}} \, ds_6^2 \,,
}
where 
\eqn\Hnought{H_0 ~\equiv~ \Om^{-4} ~=~ {m \over u^2}  \,,}
for some constant, $m$.   For $a =0$, the metric, $ds_6^2$, is the 
flat Euclidean metric on $\IR^6$, and the solution is  the standard  ``harmonic'' 
form for  $D3$-branes spread over the two-dimensional $v$-plane.  For $a \ne 0$
the metric, $ds_6^2$, is not flat, but has a curvature singularity at $v^2 = a$.  The
dilaton/axion background is also non-zero, and as pointed out in 
\GowdigereUK, the corresponding eight-metric in $F$-theory must be
hyper-K\"ahler.  

\newsec{Final comments}

We have managed to push beyond the ``harmonic rule'' for brane configurations
and obtain a solution with softly-broken supersymmetry.     In particular we
have shown how to construct a IIB supergravity solution that is the holographic
dual of $\cN=4$ Yang-Mills theory, softly broken to $\cN=2$ Yang-Mills,   
at an arbitrary, $SO(2)$-invariant point on the Coulomb branch of the $\cN=2$
theory.  More generally, we have found a natural  method for constructing 
supersymmetric solutions in the presence of multiple independent
$RR$ fluxes.   This method is based upon defining the Killing
spinors in terms of projectors that are algebraic in the metric Ansatz.
The overall Killing structure can then be used to constrain the projectors
and fix the spinor normalization.

The only complication in our procedure here is the fact that the crucial
underlying function, $c$, must satisfy a {\it non-linear} differential equation.  However,
for the metric to be asymptotically $AdS_5 \times S^5$, one must have 
$c \to 1$ at infinity.  Moreover, one can then make a perturbative expansions,
$c = 1 + \sum_n c_n$.  One then finds that $c_1$ must satisfy the linearized form
of \theeqs:
\eqn\linlapl{
{\partial\over\partial u}\,\left({v^3\over u}\,
{\partial c_1 \over\partial u}\right) ~+~
{\partial\over\partial v}\,\left({v^3\over u}\, 
{\partial c_1\over\partial v}\right) ~=~ 0\,.
}
Each of the functions, $c_n$, then satisfies the same linear equation, but 
with a source  that is quadratic in the $c_j$,  $j <n$, and their derivatives.  This
perturbation series is essentially ``seeded'' by, $c_1$, a homogeneous
solution to a second order, linear, elliptic equation in two variables.
The function, $c_1$, is, in turn, defined by some distribution of sources,
$\rho(v)$,  that is assumed to lie at $u=0$.  Whether or not this 
perturbation series is a practical solution remains to be seen, however it
does show that we have, in principle, solved the problem of 
$\cN=2$ supersymmetric flows with a rotationally symmetric 
distribution of $D3$-branes in the $(v_1, v_2)$ directions.

We suspect that the techniques employed here could be used to solve a 
variety of flow problems, and, more generally, find supersymmetric backgrounds
that involve several independent fluxes.   Work on this is continuing.

\bigskip
\leftline{\bf Acknowledgements}

This work was supported in part by funds
provided by the DOE under grant number DE-FG03-84ER-40168.

\vfill\eject
\appendix{A}{Some $\gamma$-matrix conventions and identities}

We use the mostly minus convention and label the frames and
$\gamma$-matrices from 1 to 10, with 1 being the time-like
direction. Otherwise our conventions are the same as in \SchwarzQR.
In particular, we use a pure imaginary representation of
$\gamma$-matrices,
\eqn\gareal{
(\ga^M)^*\eql -\ga^M\,,\qquad M=1,\ldots,10\,.
}
which then have the following symmetry properties:
\eqn\gasym{
(\ga^1)^T\eql -\ga^1\,,\qquad
(\ga^M)^T\eql \ga^M\,,\quad M>1\,.
}
The matrix,
\eqn\gaeleven{
\ga^{11}\eql \ga^1\ga^2\ldots\ga^{10}\,,
} is real and symmetric.  The Dirac conjugate of a spinor $\eta$ is $\overline
\eta=\eta^\dagger\ga^1$. The charge conjugation matrix is
$C=\ga^1$. Then
\eqn\symprg{
C\ga^{M_1}\ldots\ga^{M_p}\eql \cases { \hbox{symmetric for } 
p= 1,2,5,6,9,10\,.\cr
\hbox{antisymmetric for } p= 0,3,4,7,8\,.\cr}}

The spinor, $\ep_0$, which parametrizes the unbroken supersymmetries,
is defined by
\eqn\gaone{
\ga^{11}\,\ep_0\eql-\ep_0\,,\qquad
\ga^1\ga^2\ga^3\ga^4\,\ep_0\eql i\,\ep_0\,,\qquad 
\ga^6\ga^{10}\,\ep_0\eql -i\,\ep_0\,.}
Using hermiticity properties of $\ga$ matrices we then find
\eqn\gathree{
\ep_0^\dagger\,\ga^{11}\eql -\ep_0^\dagger\,,\qquad 
\ep_0^\dagger\,\ga^1\ga^2\ga^3\ga^4\eql i\,\ep_0^\dagger\,,\qquad
\ep_0^\dagger\,\ga^6\ga^{10}\eql -i\,\ep_0^\dagger\,,
}
and similar relations for $\ep_0^T$. In the notation of Section 3 
we can rewrite \gathree\ as identities for the conjugate spinor
$\overline\ep_0=\ep_0^\dagger\ga^1$,
\eqn\moreusident{
\overline\ep_0\,(1-\Pi_{0}^{(1)})\eql \overline\ep_0\,,\qquad
\overline\ep_0\,\Pi_{0}^{(2)}\eql \overline\ep_0\,.
}
In particular the first identity implies that $\overline\ep_0\ep_0=0$.

Since $(\ga^{11})^2=1$, we find that for any product of odd 
number of $\ga$-matrices
\eqn\vanisid{
\ep_0^T\,\ga^{M_1}\ldots\ga^{M_{2n+1}}\,\ep_0\eql 0\,,\qquad 
\ep_0^\dagger\,\ga^{M_1}\ldots\ga^{M_{2n+1}}\,\ep_0\,,\qquad
M_i\eql 1,\ldots,10\,.
} 
Similarly,  using $(\ga^1\ga^2\ga^3\ga^4)^2=-1$ and $(\ga^6\ga^{10})^2=-1$,
together with \gaone\ and  \gathree,
we obtain additional vanishing relations, such as 
\eqn\vanishodd{
\ep_0^T\,\ga^{M_1}\ldots\ga^{M_{2n}}\,\ep_0\eql 0\,,\qquad M_i=5,\ldots,10\,.
} 
Further vanishing relations follow from symmetry combined with the
statistics of $\ep_0$.

\vfill\eject

\appendix{B}{Killing vectors}
Here we  outline   a proof of the general result used frequently
throughout the text:
\medskip
\centerline{\vbox{ \hsize=4.5 in
\noindent
{\it Let $\epsilon$ and $\eta$ be two commuting 
Killing spinors satisfying 
\susytrpsi\ and \susytrla. Then
$${
(K^M)\eql \bar\eta\,\ga^M\,\ep+\bar\ep\ga^M\,\eta\,,
}
$$
is a Killing vector of the metric.
}}}
\medskip

The proof uses symmetry properties of $\ga$-matrices and is
essentially identical with the one in \GauntlettFZ\ for
eleven-dimensional supergravity. We show directly that the Killing
vector equation
\eqn\killeqs{
D_MK_N+D_NK_M\eql0\,,
}
is satisfied. First  rewrite \susytrpsi\ in the form:
\eqn\splitsusy{
D_M\ep\eql -{i\over 480}\, \slashed F_{(5)}\,\ga_M\,\ep-{1\over 96}\,\left(
\ga_M\ga^{PQR}-12\,\de_M{}^P\ga^{QR}\right)\,G_{PQR}\,\ep^*
\,,
}
where $\slashed F_{(5)}=F_{PQRST}\ga^{PQRST}$. Conjugating this equation we
obtain
\eqn\conjtr{
D_M\overline\ep\eql {i\over 480}\,\overline\ep\,\ga_M\slashed F_{(5)}
-{1\over 96}\,\overline{\ep^*}\,\left(\ga^{PQR}\ga_M
\pm 12 \de_M{}^P\ga^{QR}\right)\,G_{PQR}^*\,,
}
where
\eqn\covonbar{
D_M\overline\ep\eql\partial_M\overline\ep-{1\over 4}\,\overline\ep
\ga^{PQ}\,\om_{MPQ}+{i\over 2}\overline\ep\,Q_M\,.
}
Now, the left hand side of \killeqs\ becomes
\eqn\expaneqs{
(D_M\overline\eta)\,\ga_N\,\ep+\overline\eta\,\ga_N\,(D_M\ep)+
(M\leftrightarrow N)+(\eta\leftrightarrow\ep)\,.
}
Substituting \splitsusy\ and \conjtr\ in \expaneqs, we find that, 
upon symmetrization over $M$ and $N$, the terms with $F_{(5)}$ vanish, 
while in the  terms with $G_{(3)}$ only products of 
three $\gamma$-matrices remain. Those terms are of the form
\eqn\whatsleft{
\eta^T\,C\,\ga^{PQR}\,\ep\,g_{MN}\,G_{PQR}\qquad
\hbox{and}\qquad \eta^T\,C\,\ga^{PQ}_{(M}\,\ep\,G_{N)PQ}\,,
} 
and the complex conjugates. Using \symprg, we conclude that they vanish
after symmetrization in $\ep$ and $\eta$.

\vfill\eject
\appendix{C}{The supersymmetry equations}
In this appendix we list some of the equations resulting from the
substituion of the Ansatz in Section 4  in
\killspeqs\ and  indicate the variations which have given rise to each
of them. One should note that most of the equations differ from their
original form resulting from \killspeqs\ as we systematically
eliminate the $G_{u7\,10}$ and $G_{v7\,10}$ components of the
three-index tensor field and the $F_{1234u}$ and $F_{1234v}$
components of the five-index tensor field using the relations
\eqn\eqee{\eqalign{
G_{u7\,10}&\eql-i\,G_{567}+\cos(\be)\left(G_{89\,10}-i\,G_{v89}\right)\,,\cr
G_{v7\,10}&\eql i\cos(\be)\,G_{u89}\,,\cr}
}
and 
\eqn\eqac{\eqalign{
F_{1234u}&\eql {i\over 4}\csc(\be)\,G_{567}+{i\over
4}\cot(\be)\,G_{v89}\,,\cr F_{1234v}&\eql-{1\over
4}\cot(\be)\,G_{u89}\,,\cr} } which follow from $\de\lambda=0$ and
$\delta\psi_1=0$, respectively.  Throughout this appendix, after
evaluating derivatives, we  set $\phi=0$.

There is one further algebraic equation from $\delta\psi_8=0$:
\eqn\eqca{
{V_3\over V_4^2}\eql-{i\over 2}
\,\csc(\be)\,G_{567} -{1\over 2} \cot(\be)\,G_{89\,10}\,.
}
The remining equations involve derivatives of the functions in the Ansatz.
\smallskip
\item{(a)} Form $\delta\lambda=0$:
\eqn\eqaa{\eqalign{
P_u&\eql -{1\over 4} \sin(\beta)\,(G_{89\,10}-i G_{v89})\,,\cr
P_v-iP_{10}&\eql -{i\over 4}\sin(\be)\,G_{u89}\,,\cr}
}
where, we recall,
\eqn\theps{
P_{u}\eql {E_u\,B\over 1-B^2}\,,\qquad
P_{v}\eql {E_v\,B\over 1-B^2}\,,
}
and
\eqn\theqs{
P_{10}\eql {2 i\over V_5}\,{B\over 1-B^2}\,,\qquad
Q_{10}\eql -{2\over V_5}\,{B^2\over 1-B^2}\,.
}

\item{(b)} From $\delta\psi_u$ and $\delta\psi_v$:
\eqn\eqba{\eqalign{
E_u\beta&\eql -i\,G_{567}+
{1\over 2}\,\cos(\be)\left(G_{89\,10}-iG_{v89}\right)\,,\cr
E_v\beta&\eql {i\over 2}\,\cos(\be)\,G_{u89}\,,\cr}
}
and 
\eqn\eqda{\eqalign{
E_u\log(\Om)& \eql {i\over 4}\,\cot(\be)\,G_{567}\cr
&\qquad
+{i\over 16}(3+\cos(2\be))\csc(\be)\,G_{v89}
+{1\over 8}\,\sin(\be)\,G_{89\,10}\,,\cr
E_v\,\log(\Om)&\eql-{i\over 16}(3+\cos(2\be))\csc(\be)\,G_{u89}\,,\cr}
}
\item{(c)} From $\delta\psi_7=0$:
\eqn\eqdc{\eqalign{
E_u\log V_3&\eql -{3i\over 4}\cot(\be)\,G_{567}+
{1\over 16}(5+3\cos(2\be))\csc(\be)\,G_{89\,10}\cr
&\qquad\qquad-{i\over 16} (1+3\cos(2\be))\csc(\be)\,G_{v89}\,,\cr
E_v\log V_3&\eql {i\over 16} (1+3\cos(2\be))\csc(\be) \,G_{u89}\,,\cr}
}
\item{(d)} From $\delta\psi_{8,9}=0$:
\eqn\eqdd{\eqalign{
E_u\log V_4&\eql {2\over V_4}+{i\over4}\cot(\be)\,G_{567}
-{1\over 16}(7+\cos(2\be))\csc(\be)\,G_{89\,10}\cr
&\qquad -{i\over 16}(5-\cos(2\be))\csc(\be)\,G_{v89}
\cr
E_v\log V_4&\eql {i\over 16} (5-\cos(2\be))\,\csc(\beta)\,G_{u89}\cr}
}
\item{(e)} From $\delta\psi_{10}=0$:
\eqn\eqdb{\eqalign{
E_u\log V_5&\eql {E_u B\over 1-B^2}-2 E_u(\Om^{1/2})\cr
E_v\log V_5&\eql {E_vB\over 1-B^2}+{1\over V_5}\left({1+B\over1-B}\right)+
{i\over 16}(3+\cos(2\be))\csc(\be)\,G_{u89}\cr}
}

\noindent
After passing to the new new coordinates $u$ and $v$, we 
obtain equations for $V_u$ and $V_v$.
\smallskip

\item{(f)} From $\delta\psi_{u}=0$ and $\delta\psi_{v}=0$:
\eqn\eqee{\eqalign{
E_u\log V_v &\eql -{i\over 4}\cot(\be)\, G_{567}
-{i\over 16}(5-\cos(2\be))\csc(\be)\, G_{v89}\cr
&\qquad
+{1\over 8} \sin(\be)\,G_{89\,10}\,,\cr
E_v\log V_u &\eql {i\over 16}(5-\cos(2\be))\csc(\be)\, G_{u89}\,.\cr
}
}

\listrefs

\vfill\eject\end